\documentclass[12pt]{article}

\usepackage{amsfonts, amsmath}

\newcommand{\be}{\begin{equation}} \newcommand{\ee}{\end{equation}}

\begin{document}
\title{Deformed Density Matrix and Generalized Uncertainty Relation
 in  Thermodynamics
} \thispagestyle{empty}

\author{A.E.Shalyt-Margolin\thanks
{Fax: (+375) 172 326075; e-mail: a.shalyt@mail.ru;
alexm@hep.by}\hspace{10pt} and \hspace{5pt} A.Ya.Tregubovich
\thanks{Phone (+375) 172 840441; e-mail a.tregub@open.by}
}
\date{}
\maketitle
 \vspace{-15pt}
{\footnotesize\noindent {\large $^*$} National Centre of High Energy and
Particle Physics Bogdanovich Str.153, Minsk \hspace*{8pt} 220040, Belarus\\
{\large $^\dagger$} Institute of Physics National Academy of Sciences
                   Skoryna av.68, Minsk\\\hspace*{8pt} 220072, Belarus}\\

{\bf\small\noindent Abstract}\\ {\footnotesize
A generalization of
the thermodynamic uncertainty relations is proposed. It is done by
introducing of an additional term proportional to the interior
energy  into the standard thermodynamic uncertainty relation that
leads to existence  of the lower limit of inverse
temperature. The authors are of the opinion that the approach proposed
may lead to proof of these relations. To this end, the statistical mechanics
deformation at Planck scale. The statistical mechanics deformation
is constructed by analogy to the earlier quantum mechanical
results. As previously, the primary object is a density matrix,
but now the statistical one. The obtained deformed object is
referred to as a statistical density pro-matrix. This object is
explicitly described, and it is demonstrated that there is a
complete analogy in the construction and properties of quantum
mechanics and statistical density matrices at Plank scale (i.e.
density pro-matrices). It is shown that an ordinary statistical
density matrix occurs in the low-temperature limit at temperatures
much lower than the Plank's. The associated deformation of a
canonical Gibbs distribution is given explicitly.}
 \vspace{0.5cm}
{\ttfamily{\footnotesize
\\ PACS: 03.65;05.20;05.70\\ \noindent Keywords:
                   generalized   uncertainty relations; generalized
                   uncertainty\\ relations in thermodynamics;deformed density matixes}}

\rm\normalsize \vspace{0.5cm}

\section{Introduction}
In this paper generalization of the thermodynamic uncertainty
relations is proposed. It is done by introducing of an additional
term proportional to the interior energy  into the standard
thermodynamic uncertainty relation that leads to existence  of the
lower limit of inverse temperature. Consequently, statistical mechanics
at Planck scale should be deformed. As is known,
at Planck scale Quantum Mechanics (QM) undergoes variation:
it should be subjected to deformation also. This is realized due
to the presence of the Generalized Uncertainty Relations (GUR) and
hence the fundamental length \cite{r1},\cite{r2}. The deformation
in Quantum Mechanics at Planck scale takes different paths:
commutator deformation (Heisenberg's algebra deformation)
\cite{r4},\cite{r5} or density matrix deformation \cite{r7},
\cite{r8}. In the present work the second approach is extended by
the authors to the Statistical Mechanics at Plank scale. To this
end, a deformed statistical density matrix, also called a
statistical density pro-matrix, is constructed as a complete
analog to the deformed quantum mechanics matrix. In Quantum
Mechanics with fundamental length (QMFL) the deformation parameter
was represented by the value $\alpha=l_{min}^{2}/x^{2}$ where $x$
is the scale, whereas in case of the Statistical Mechanics this
value will be $\tau = T^{2}/T^{2}_{max}$ where $T_{max}$ is a
maximum temperature of the order of the Planck's. Existence of
$T_{max}$ follows from (GUR) for the "energy - time" pair. The
limitations on the parameter variation interval are the same. In
this way it is demonstrated that there exists a complete analogy
in the construction and properties of quantum mechanics and
statistical density matrices at Planck scale (density
pro-matrices). It should be noted that an ordinary statistical
density matrix appears in the low-temperature limit (at
temperatures much lower than the Planck's).  The associated
deformation of a canonical Gibbs distribution is described
explicitly.

\section{Generalized Uncertainty Relation in \\Thermodynamics}

 It is well known that in thermodynamics an inequality for the pair interior energy -
 inverse temperature, which is completely analogous to the standard uncertainty
 relation in quantum mechanics \cite{r9} can be written down \cite{r10} -- \cite{r12}. The
 only (but essential) difference of this inequality from the quantum mechanical
 one is that the main quadratic fluctuation is defined by means of
 classical partition function rather than by quantum mechanical expectation values.
 In the last 14 - 15 years a lot of papers appeared in which the usual
 momentum-coordinate uncertainty relation has been modified at very high
 energies of order Planck energy $E_p$ \cite{r1}--\cite{castro1}. In this note we
 propose simple reasons for modifying the thermodynamic uncertainty relation at
 Planck energies. This modification results in existence of the minimal
 possible main quadratic fluctuation of the inverse temperature. Of course we
 assume that all the thermodynamic quantities used are properly defined so that
 they have physical sense at such high energies.

We start with usual Heisenberg uncertainty relations \cite{r9} for
momentum - coordinate:
\begin{equation}\label{U1}
 \Delta x\geq\frac{\hbar}{\Delta p}.
\end{equation}
 It was shown that at the Planck
 scale a high-energy term must appear:
\begin{equation}\label{U2}
\Delta x\geq\frac{\hbar}{\Delta p}+ \alpha^{\prime}
L_{p}^2\frac{\triangle p}{\hbar}
\end{equation}
where $L_{p}$ is the Planck length $L_{p}^2 = G\hbar /c^3 \simeq
1,6\;10^{-35}m$ and $\alpha^{\prime}$ is a constant. In \cite{r3}
this term is derived from the string theory, in \cite{r1}
 it follows from the simple estimates of Newtonian gravity and quantum mechanics,
 in \cite{r4} it comes from the black hole physics, other methods can also be
 used \cite{r5},\cite{r6}.
Relation (\ref{U2}) is quadratic in $\Delta p$
\begin{equation}\label{U4}
\alpha^{\prime} L_{p}^2\, ({\Delta p})^2 - \hbar\,\Delta x\Delta p
+ \hbar^2 \leq0
\end{equation}
 and therefore leads to the fundamental length
\begin{equation}\label{U5}
 \Delta x_{min}=2 \surd \alpha^{\prime} L_{p}
\end{equation}
  Using relations (\ref{U2}) it is easy to obtain a similar relation for the
 energy - time pair. Indeed (\ref{U2}) gives
\begin{equation}\label{U6}
\frac{\Delta x}{c}\geq\frac{\hbar}{\Delta p c }+\alpha^{\prime}
L_{p}^2\,\frac{\Delta p}{c \hbar},
\end{equation}
then
\begin{equation}\label{U7}
\Delta t\geq\frac{\hbar}{\Delta
E}+\alpha^{\prime}\frac{L_{p}^2}{c^2}\,\frac{\Delta p
c}{\hbar}=\frac{\hbar}{\Delta
E}+\alpha^{\prime}t_{p}^2\,\frac{\Delta E}{ \hbar}.
\end{equation}
where the smallness of $L_p$ is taken into account so that the difference
between $\Delta E$ and $\Delta (pc)$ can be neglected and $t_{p}$  is the
Planck time $t_{p}=L_p/c=\sqrt{G\hbar/c^5}\simeq 0,54\;10^{-43}sec$.
Inequality (\ref{U7}) gives analogously to (\ref{U2}) the lower boundary
for time $\Delta t\geq2t_{p}$ determining the fundamental time
\begin{equation}\label{U10b}
 \Delta t_{min}=2\sqrt{\alpha^{\prime}}t_{p}
 \end{equation}
 Thus, the inequalities discussed can be rewritten in a standard form
\begin{equation}\label{U11b}
\left\{ \begin{array}{ll} \Delta x &
\geq\frac{\displaystyle\hbar}{\displaystyle\Delta
p}+\alpha^{\prime} \left(\frac{\displaystyle\Delta
p}{\displaystyle P_{pl}}\right)\,
\frac{\displaystyle\hbar}{\displaystyle P_{pl}}
\\
 & \\
 \Delta t & \geq\frac{\displaystyle\hbar}{\displaystyle\Delta E}+\alpha^{\prime}
 \left(\frac{\displaystyle\Delta E}{\displaystyle E_{p}}\right)\,
 \frac{\displaystyle\hbar}{\displaystyle E_{p}}
\end{array} \right.
\end{equation}
where $P_{pl}=E_p/c=\sqrt{\hbar c^3/G}$.
 Now we
consider the thermodynamics uncertainty relations between the inverse temperature
and interior energy of a macroscopic ensemble
\begin{equation}\label{U12}
\Delta \frac{1}{T}\geq\frac{k}{\Delta U}.
\end{equation}
where $k$ is the Boltzmann constant. \\ N.Bohr \cite{r10} and
W.Heisenberg \cite{r11} first pointed out that such kind of
uncertainty principle should take place in thermodynamics. The
thermodynamic uncertainty  relations (\ref{U12})  were proved by
many authors and in various ways \cite{r12}. Therefore their
validity does not raise any doubts. Nevertheless, relation
(\ref{U12}) was proved in view of the standard model of the
infinite-capacity heat bath encompassing the ensemble. But it is
obvious from the above inequalities that at very high energies the
capacity of the heat bath can no longer to be assumed infinite at
the Planck scale. Indeed, the total energy of the pair heat bath -
ensemble may be arbitrary large but finite merely as the universe
is born at a finite energy. Hence the quantity that can be
interpreted as the temperature of the ensemble must have the upper
limit and so does its main quadratic deviation. In other words the
quantity $\Delta (1/T)$ must be bounded from below. But in this
case an additional term should be introduced into (\ref{U12})
\begin{equation}\label{U12a}
\Delta \frac{1}{T}\geq\frac{k}{\Delta U} + \eta\,\Delta U
\end{equation}
where $\eta$ is a coefficient. Dimension and symmetry reasons give
$$ \eta \sim \frac{k}{E_p^2}\enskip or\enskip \eta =
\alpha^{\prime} \frac{k}{E_p^2} $$
 As in the previous cases
inequality (\ref{U12a}) leads to the fundamental (inverse)
temperature.
\begin{equation}\label{U15}
T_{max}=\frac{\hbar}{2\surd \alpha^{\prime}t_{p}
k}=\frac{\hbar}{\Delta t_{min} k}, \quad \beta_{min} = {1\over
kT_{max}} =  \frac{\Delta t_{min}}{\hbar}
\end{equation}
It should be noted that the same conclusion about the existence of
the maximal temperature in Nature can be made also considering
black hole evaporation \cite{castro2}.
 \\ Thus, we obtain the
system of generalized uncertainty relations in a symmetric form
\begin{equation}\label{U17}
\left\{
\begin{array}{lll}
\Delta x & \geq & \frac{\displaystyle\hbar}{\displaystyle\Delta
p}+ \alpha^{\prime} \left(\frac{\displaystyle\Delta
p}{\displaystyle P_{pl}}\right)\,
\frac{\displaystyle\hbar}{\displaystyle P_{pl}}+... \\ &  &  \\
\Delta t & \geq & \frac{\displaystyle\hbar}{\displaystyle\Delta
E}+\alpha^{\prime} \left(\frac{\displaystyle\Delta
E}{\displaystyle E_{p}}\right)\,
\frac{\displaystyle\hbar}{\displaystyle E_{p}}+...\\
  &  &  \\
  \Delta \frac{\displaystyle 1}{\displaystyle T}& \geq &
  \frac{\displaystyle k}{\displaystyle\Delta U}+\alpha^{\prime}
  \left(\frac{\displaystyle\Delta U}{\displaystyle E_{p}}\right)\,
  \frac{\displaystyle k}{\displaystyle E_{p}}+...
\end{array} \right.
\end{equation}
or in the equivalent form
\begin{equation}\label{U18}
\left\{
\begin{array}{lll}
\Delta x & \geq & \frac{\displaystyle\hbar}{\displaystyle\Delta
p}+\alpha^{\prime} L_{p}^2\,\frac{\displaystyle\Delta
p}{\displaystyle\hbar}+... \\
  &  &  \\
  \Delta t & \geq &  \frac{\displaystyle\hbar}{\displaystyle\Delta E}+\alpha^{\prime}
  t_{p}^2\,\frac{\displaystyle\Delta E}{\displaystyle\hbar}+... \\
  &  &  \\

  \Delta \frac{\displaystyle 1}{\displaystyle T} & \geq &
  \frac{\displaystyle k}{\displaystyle\Delta U}+\alpha^{\prime}
  \frac{\displaystyle 1}{\displaystyle T_{p}^2}\,
  \frac{\displaystyle\Delta U}{\displaystyle k}+...
\end{array} \right.
\end{equation}
where the dots mean the existence of higher order corrections as
in \cite{r21}.
 Here $T_{p}$ is the Planck temperature:
$T_{p}=\frac{E_{p}}{k}$.
\\ In the conclusion of this section we would like to note that the restriction on
the heat bath made above turns the equilibrium   partition
function to be non-Gibbsian \cite{r13}.
\\ Note that the last inequality is symmetrical to the second one
with respect to the substitution \cite{r15}
\\
$$ t\mapsto\frac{1}{T}, \hbar\mapsto k,\triangle E\mapsto
\triangle U . $$ However this observation can by no means be
regarded as a rigorous proof of the generalized uncertainty
relation in thermodynamics.
\\
There is a reason to believe that a rigorous justification for the
last (thermodynamic) inequalities in systems (\ref{U17}) and
(\ref{U18}) may be made by means of a certain deformation of Gibbs
distribution.
\\
Let us outline the main aspects of above-considered deformation.
In our opinion it could be obtained as the result of
density-matrix deformation in Statistical Mechanics (see
\cite{r16}, Section 2, Paragraph 3):

\begin{equation}\label{U19}
\rho=\sum_{n}\omega_{n}|\varphi_{n}><\varphi_{n}|,
\end{equation}
where probability is given by
\\
$$\omega_{n}=\frac{1}{Q}\exp(-\beta E_{n}).$$
\\
Deformation of density matrix $\rho$ (\ref{U19}) can be carried
out similarly to deformation of density matrix (density
pro-matrix) in Quantum Mechanics at Planck's scale (see
\cite{r7},\cite{r8}). Proceeding with this analogy density matrix
$\rho$ in (\ref{U19})
 should be changed by $\rho(\tau)$, where $\tau$ is a parameter of deformation.
Deformed density matrix  must fulfill the condition
$\rho(\tau)\approx\rho$ when $T\ll T_{p}$. By analogy with
\cite{r7},\cite{r8}, only probabilities $\omega_{n}$ are subject
of deformation in (\ref{U19}), changing by $\omega_{n}(\tau)$ and
correspondingly deformed statistical density matrix is
\begin{equation}\label{U20}
\rho(\tau)=\sum_{n}\omega_{n}(\tau)|\varphi_{n}><\varphi_{n}|.
\end{equation}

 This approach in our
opinion could give us the possibility to obtain Deformed Canonical
Distribution as well as a rigorous proof of thermodynamical
general uncertainty relations. In section 4 the construction of
such a deformed statistical mechanics at Planck scale
is demonstrated. However, first it seems expedient to outline briefly
the principal features of the corresponding deformation in QM.

\section {Deformation of Quantum-Mechanics Density Matrix at Planck
Scale}

In this section the principal features of QMFL construction with
the
 use of the density matrix deformation are briefly outlined
 \cite{r8}.

  As mentioned above, for the fundamental deformation parameter
we use $\alpha = l_{min}^{2 }/x^{2 }$ where $x$ is the scale. In
contrast with \cite{r8}, for the deformation parameter we use
$\alpha$ rather than  $\beta$ to avoid confusion, since quite a
distinct value is denoted by $\beta$ in Statistical
Mechanics:$\beta=1/kT$.
\\
\noindent {\bf Definition 1.} {\bf(Quantum Mechanics with
Fundamental Length)}
\\
\noindent Any system in QMFL is described by a density pro-matrix
of the form $${\bf
\rho(\alpha)=\sum_{i}\omega_{i}(\alpha)|i><i|},$$ where
\begin{enumerate}
\item $0<\alpha\leq1/4$;
\item The vectors $|i>$ form a full orthonormal system;
\item $\omega_{i}(\alpha)\geq 0$ and for all $i$  the
finite limit $\lim\limits_{\alpha\rightarrow
0}\omega_{i}(\alpha)=\omega_{i}$ exists;
\item
$Sp[\rho(\alpha)]=\sum_{i}\omega_{i}(\alpha)<1$,
$\sum_{i}\omega_{i}=1$;
\item For every operator $B$ and any $\alpha$ there is a
mean operator $B$ depending on  $\alpha$:\\
$$<B>_{\alpha}=\sum_{i}\omega_{i}(\alpha)<i|B|i>.$$

\end{enumerate}
Finally, in order that our definition 1 agree with the result of
section 2, the following condition must be fulfilled:
\begin{equation}\label{U1b}
Sp[\rho(\alpha)]-Sp^{2}[\rho(\alpha)]\approx\alpha.
\end{equation}
Hence we can find the value for $Sp[\rho(\alpha)]$ satisfying the
condition of definition 1:
\begin{equation}\label{U2b}
Sp[\rho(\alpha)]\approx\frac{1}{2}+\sqrt{\frac{1}{4}-\alpha}.
\end{equation}

According to point 5),  $<1>_{\alpha}=Sp[\rho(\alpha)]$. Therefore
for any scalar quantity $f$ we have $<f>_{\alpha}=f
Sp[\rho(\alpha)]$. In particular, the mean value
$<[x_{\mu},p_{\nu}]>_{\alpha}$ is equal to
\\
$$<[x_{\mu},p_{\nu}]>_{\alpha}= i\hbar\delta_{\mu,\nu}
Sp[\rho(\alpha)]$$
\\
We denote the limit $\lim\limits_{\alpha\rightarrow
0}\rho(\alpha)=\rho$ as the density matrix. Evidently, in the
limit $\alpha\rightarrow 0$ we return to QM.

As follows from definition 1,
$<(j><j)>_{\alpha}=\omega_{j}(\alpha)$, from whence the
completeness condition by $\alpha$ is
\\$<(\sum_{i}|i><i|)>_{\alpha}=<1>_{\alpha}=Sp[\rho(\alpha)]$. The
norm of any vector $|\psi>$ assigned to  $\alpha$ can be defined
as
\\$<\psi|\psi>_{\alpha}=<\psi|(\sum_{i}|i><i|)_{\alpha}|\psi>
=<\psi|(1)_{\alpha}|\psi>=<\psi|\psi> Sp[\rho(\alpha)]$, where
$<\psi|\psi>$ is the norm in QM, i.e. for $\alpha\rightarrow 0$.
Similarly, the described theory may be interpreted using a
probabilistic approach, however requiring  replacement of $\rho$
by $\rho(\alpha)$ in all formulae.

\renewcommand{\theenumi}{\Roman{enumi}}
\renewcommand{\labelenumi}{\theenumi.}
\renewcommand{\labelenumii}{\theenumii.}

It should be noted:

\begin{enumerate}
\item The above limit covers both Quantum
and Classical Mechanics. Indeed, since $\alpha\sim L_{p}^{2 }/x^{2
}=G \hbar/c^3 x^{2}$, we obtain:
\begin{enumerate}
\item $(\hbar \neq 0,x\rightarrow
\infty)\Rightarrow(\alpha\rightarrow 0)$ for QM;
\item $(\hbar\rightarrow 0,x\rightarrow
\infty)\Rightarrow(\alpha\rightarrow 0)$ for Classical Mechanics;
\end{enumerate}
\item As a matter of fact, the deformation parameter $\alpha$
should assume the value $0<\alpha\leq1$.  However, as seen from
(\ref{U2b}), $Sp[\rho(\alpha)]$ is well defined only for
$0<\alpha\leq1/4$, i.e. for $x=il_{min}$ and $i\geq 2$ we have no
problems at all. At the point, where $x=l_{min}$, there is a
singularity related to complex values assumed by
$Sp[\rho(\alpha)]$ , i.e. to the impossibility of obtaining a
diagonalized density pro-matrix at this point over the field of
real numbers. For this reason definition 1 has no sense at the
point $x=l_{min}$.

\item We consider possible solutions for (\ref{U1}).
For instance, one of the solutions of (\ref{U1}), at least to the
first order in $\alpha$, is $$\rho^{*}(\alpha)=\sum_{i}\alpha_{i}
exp(-\alpha)|i><i|,$$ where all $\alpha_{i}>0$ are independent of
$\alpha$  and their sum is equal to 1. In this way
$Sp[\rho^{*}(\alpha)]=exp(-\alpha)$. Indeed, we can easily verify
that \begin{equation}\label{U3}
Sp[\rho^{*}(\alpha)]-Sp^{2}[\rho^{*}(\alpha)]=\alpha+O(\alpha^{2}).
\end{equation}
 Note that in the momentum representation $\alpha\sim p^{2}/p^{2}_{pl}$,
where $p_{pl}$ is the Planck momentum. When present in matrix
elements, $exp(-\alpha)$ can damp the contribution of great
momenta in a perturbation theory.
\item It is clear that within the proposed description the
states with a unit probability, i.e. pure states, can appear only
in the limit $\alpha\rightarrow 0$, when all $\omega_{i}(\alpha)$
except for one are equal to zero or when they tend to zero at this
limit. In our treatment pure state are states, which can be
represented in the form $|\psi><\psi|$, where $<\psi|\psi>=1$.

\item We suppose that all the definitions concerning a
density matrix can be transferred to the above-mentioned
deformation of Quantum Mechanics (QMFL) through changing the
density matrix $\rho$ by the density pro-matrix $\rho(\alpha)$ and
subsequent passage to the low energy limit $\alpha\rightarrow 0$.
Specifically, for statistical entropy we have
\begin{equation}\label{U4b}
S_{\alpha}=-Sp[\rho(\alpha)\ln(\rho(\alpha))].
\end{equation}
The quantity of $S_{\alpha}$ seems never to be equal to zero as
$\ln(\rho(\alpha))\neq 0$ and hence $S_{\alpha}$ may be equal to
zero at the limit $\alpha\rightarrow 0$ only.
\end{enumerate}
Some Implications:
\begin{enumerate}
\item If we carry out measurement on the pre-determined scale, it is
impossible to regard the density pro-matrix as a density matrix
with an accuracy better than particular limit $\sim10^{-66+2n}$,
where $10^{-n}$ is the measuring scale. In the majority of known
cases this is sufficient to consider the density pro-matrix as a
density matrix. But on Planck's scale, where the quantum
gravitational effects and Plank energy levels cannot be neglected,
the difference between $\rho(\alpha)$ and  $\rho$ should be taken
into consideration.

\item Proceeding from the above, on Planck's scale the
notion of Wave Function of the Universe (as introduced in
\cite{r17}) has no sense, and quantum gravitation effects in this
case should be described with the help of density pro-matrix
$\rho(\alpha)$ only.
\item Since density pro-matrix $\rho(\alpha)$ depends on the measuring
scale, evolution of the Universe within the inflation model
paradigm \cite{r18} is not a unitary process, or otherwise the
probabilities $p_{i}=\omega_{i}(\alpha)$  would be preserved.
\end{enumerate}
\section {Deformation of Statistical Density Matrix}
 It follows that we have a maximum energy of the order
 of Planck's from an inequality (\ref{U7}):
\\
$$E_{max}\sim E_{p}$$
\\
Proceeding to the Statistical Mechanics, we further assume that an
internal energy of any ensemble U could not be in excess of
$E_{max}$ and hence temperature $T$ could not be in excess of
$T_{max}=E_{max}/k \sim T_{p}$. Let us consider density matrix in
Statistical Mechanics :
\begin{equation}\label{U8}
\rho_{stat}=\sum_{n}\omega_{n}|\varphi_{n}><\varphi_{n}|,
\end{equation}
where the probabilities are given by
\\
$$\omega_{n}=\frac{1}{Q}\exp(-\beta E_{n})$$ and
\\
$$Q=\sum_{n}\exp(-\beta E_{n})$$
\\
Then for a canonical Gibbs ensemble the value
\begin{equation}\label{U9}
\overline{\Delta(1/T)^{2}}=Sp[\rho_{stat}(\frac{1}{T})^{2}]
-Sp^{2}[\rho_{stat}(\frac{1}{T})],
\end{equation}
is always equal to zero, and this follows from the fact that
$Sp[\rho_{stat}]=1$. However, for very high temperatures $T\gg0$
we have $\Delta (1/T)^{2}\approx 1/T^{2}\geq 1/T_{max}^{2}$. Thus,
for $T\gg0$ a statistical density matrix $\rho_{stat}$ should be
deformed so that in the general case
\begin{equation}\label{U10}
Sp[\rho_{stat}(\frac{1}{T})^{2}]-Sp^{2}[\rho_{stat}(\frac{1}{T})]
\approx \frac{1}{T_{max}^{2}},
\end{equation}
or \begin{equation}\label{U11} Sp[\rho_{stat}]-Sp^{2}[\rho_{stat}]
\approx \frac{T^{2}}{T_{max}^{2}},
\end{equation}
In this way $\rho_{stat}$ at very high $T\gg 0$ becomes dependent
on the parameter $\tau = T^{2}/T_{max}^{2}$, i.e. in the most
general case
\\
$$\rho_{stat}=\rho_{stat}(\tau)$$ and $$Sp[\rho_{stat}(\tau)]<1$$
\\
and for $\tau\ll 1$ we have $\rho_{stat}(\tau)\approx\rho_{stat}$
(formula (\ref{U8})) .\\ This situation is identical to the case
associated with the deformation parameter $\alpha = l_{min}^{2
}/x^{2}$ of QMFL given in section  3. That is the condition
$Sp[\rho_{stat}(\tau)]<1$ has an apparent physical meaning when:
\begin{enumerate}
 \item At temperatures close to $T_{max}$ some portion of information
about the ensemble is inaccessible in accordance with the
probability that is less than unity, i.e. incomplete probability.
 \item And vice versa, the longer is the distance from $T_{max}$ (i.e.
when approximating the usual temperatures), the greater is the
bulk of information and the closer is the complete probability to
unity.
\end{enumerate}
 Therefore similar to the introduction of the deformed
quantum-mechanics density matrix in section 3 of \cite{r8} and in
previous section of this paper,we give the following
\\
\noindent {\bf Definition 2.} {\bf(Deformation of Statistical
Mechanics)} \noindent \\Deformation of Gibbs distribution valid
for temperatures on the order of the Planck's $T_{p}$ is described
 by deformation of a statistical density matrix
  (statistical density pro-matrix) of the form
\\$${\bf \rho_{stat}(\tau)=\sum_{n}\omega_{n}(\tau)|\varphi_{n}><\varphi_{n}|}$$
 having the deformation parameter
$\tau = T^{2}/T_{max}^{2}$, where
\begin{enumerate}
\item $0<\tau \leq 1/4$;
\item The vectors $|\varphi_{n}>$ form a full orthonormal system;
\item $\omega_{n}(\tau)\geq 0$ and for all $n$ at $\tau \ll 1$
 we obtain
 $\omega_{n}(\tau)\approx \omega_{n}=\frac{1}{Q}\exp(-\beta E_{n})$
In particular, $\lim\limits_{T_{max}\rightarrow \infty
(\tau\rightarrow 0)}\omega_{n}(\tau)=\omega_{n}$
\item
$Sp[\rho_{stat}(\tau)]=\sum_{n}\omega_{n}(\tau)<1$,
$\sum_{n}\omega_{n}=1$;
\item For every operator $B$ and any $\tau$ there is a
mean operator $B$ depending on  $\tau$ \\
$$<B>_{\tau}=\sum_{n}\omega_{n}(\tau)<n|B|n>.$$
\end{enumerate}
Finally, in order that our Definition 2 agree with the formula
(\ref{U11}), the following condition must be fulfilled:
\begin{equation}\label{U12b}
Sp[\rho_{stat}(\tau)]-Sp^{2}[\rho_{stat}(\tau)]\approx \tau.
\end{equation}
Hence we can find the value for $Sp[\rho_{stat}(\tau)]$
 satisfying
the condition of Definition 2 (similar to Definition 1):
\begin{equation}\label{U13}
Sp[\rho_{stat}(\tau)]\approx\frac{1}{2}+\sqrt{\frac{1}{4}-\tau}.
\end{equation}
It should be noted:

\begin{enumerate}
\item The condition $\tau \ll 1$ means that $T\ll T_{max}$ either
$T_{max}=\infty$ or both in accordance with a normal Statistical
Mechanics and canonical Gibbs distribution (\ref{U8})
\item Similar to QMFL in Definition 1, where the deformation
parameter $\alpha$ should assume the value $0<\alpha\leq1/4$. As
seen from (\ref{U13}), here $Sp[\rho_{stat}(\tau)]$ is well
defined only for $0<\tau\leq1/4$. This means that the feature
occurring in QMFL at the point of the fundamental length
$x=l_{min}$ in the case under consideration is associated with the
fact that {\bf highest  measurable temperature of the ensemble is
always} ${\bf T\leq \frac{1}{2}T_{max}}$.

\item The constructed deformation contains all four fundamental constants:
 $G,\hbar,c,k$ as $T_{max}=\varsigma T_{p}$,where $\varsigma$
 is the denumerable function of  $\alpha^{\prime}$
(\ref{U2})and $T_{p}$, in its turn, contains all the
above-mentioned
 constants.

\item Again similar to QMFL, as a possible solution for (\ref{U12})
we have an exponential ansatz
\\
$$\rho_{stat}^{*}(\tau)=\sum_{n}\omega_{n}(\tau)|n><n|=\sum_{n}
exp(-\tau) \omega_{n}|n><n|$$
\\
\begin{equation}\label{U14}
Sp[\rho_{stat}^{*}(\tau)]-Sp^{2}[\rho_{stat}^{*}(\tau)]=\tau+O(\tau^{2}).
\end{equation}
In such a way with the use of an exponential ansatz (\ref{U14})
the deformation of a canonical Gibbs distribution at Planck scale
(up to factor $1/Q$) takes an elegant and completed form:
\begin{equation}\label{U15b}
{\bf \omega_{n}(\tau)=exp(-\tau)\omega_{n}= exp(-\frac{T^{2}}
{T_{max}^{2}}-\beta E_{n})}
\end{equation}
where $T_{max}= \varsigma T_{p}$

\section{Conclusion}

It has been demonstrated that a nature of deformations in Quantum
and Statistical Mechanics at Plank scale is essentially identical.
Still further studies are required to look into variations of the
formulae for entropy and other quantities in this deformed
Statistical Mechanics. Of particular interest is the problem of a
rigorous proof for the Generalized Uncertainty Relations (GUR) in
Thermodynamics (section 2 of the present paper and
\cite{r14},\cite{r19}) as a complete analog of the corresponding
relations in Quantum Mechanics \cite{r1}, \cite{r3,r4,r5,r6},
in turn necessitating the deformation of Gibbs distribution.
The present paper as an integration of
\cite{r19},\cite{r20}is aimed at the solution of this problem.

\end{enumerate}


\begin{thebibliography}{99}
%
%
\bibitem{r1}
R.J.Adler and D.I.Santiago,On Gravity and the Uncertainty
Principle, Mod.Phys.Lett.A14(1999)1371[gr-qc/9904026]
%
%
\bibitem{r2}
L.Garay,Quantum Gravity and Minimum Length
Int.J.Mod.Phys.A.A10(1995)145[gr-qc/9403008]
%
%
\bibitem{r3}
G.Veneziano,A stringly nature needs just two constant
Europhys.Lett.2(1986)199;D.Amati,M.Ciafaloni
 and G.Veneziano,Can spacetime be probed below the
 string size? Phys.Lett.B216(1989)41;
E.Witten, Reflections on the Fate of Spacetime
Phys.Today,49(1996)24
%
%
\bibitem{r4}
M.Maggiore, A Generalized Uncertainty Principle in Quantum Gravity
Phys.Lett.B304(1993)65,[hep-th/9301067]
%
%
\bibitem{r5}
M.Maggiore,Quantum Groups,Gravity and Generalized Uncertainty
Principle Phys.Rev.D49(1994)5182,[hep-th/9305163]; The algebraic
structure of the generalized uncertainty principle
Phys.Lett.B319(1993)83,[hep-th/9309034];S.Capozziello,G.Lambiase
and G.Scarpetta, The Generalized Uncertainty Principle from
Quantum Geometry [gr-qc/9910017]
%
%
\bibitem{r6}
D.V.Ahluwalia,Wave-Particle duality at the Planck scale: Freezing
of neutrino oscillations Phys.Lett. A275 (2000)31,
[gr-qc/0002005];Interface of Gravitational and Quantum Realms
Mod.Phys.Lett. A17(2002)1135,[gr-qc/0205121]
%
%
\bibitem{castro1} C. Castro Foundations of Physics 30 (2000) 1301 ;
hep-th/0001023;
C. Castro : J.Chaos, Solitons and Fractals 11
(2000) 1663 .

\bibitem{castro2} C. Castro, A. Granik Foundations of Physics vol. 33  no.3  ( 2003 ) 445;

\bibitem{r7}
A.E.Shalyt-Margolin Fundamental Length,Deformed Density Matrix and
New View on the Black Hole Information
Paradox,[gr-qc/0207074];A.E.Shalyt-Margolin and A.Ya.Tregubovich,
Generalized Uncertainty Relations,Fundamental Length and Density
Matrix,[gr-qc/0207068];A.E.Shalyt-Margolin and J.G.Suarez. Density
Matrix and Dynamical aspects of Quantum Mechanics with Fundamental
Length, [gr-qc/0211083]; A.E.Shalyt-Margolin and
J.G.Suarez,Quantum Mechanics of the Early Universe and its
Limiting Transition,[gr-qc/0302119]
%
%
\bibitem{r8}
A.E.Shalyt-Margolin and J.G.Suarez,Quantum Mechanics at Planck's
scale and Density Matrix,Intern.Journ.of Mod.Phys.D.12(2003)1265
%
%
\bibitem{r9}
W.Heisenberg,Uber den anschaulichen Inhalt der
quantentheoretischen Kinematik und Mechanik,
 Zeitsch.fur Phys,43(1927)172
%
%
\bibitem{r10}
N.Bohr, Faraday Lectures pp. 349-384, 376-377 Chemical Society,
London (1932)
%
%
\bibitem{r11}
W.Heisenberg, Der Teil und Das Ganze ch 9 R.Piper, Munchen (1969)
%
%
\bibitem{r12}
J.Lindhard Complementarity between energy and temperature. In: The
Lesson of Quantum Theory, Ed. by J. de Boer, E.Dal and O.Ulfbeck
North-Holland, Amsterdam (1986); B.Lavenda, Statistical Physics: a
Probabilistic Approach J.Wiley and Sons, N.Y. (1991);
B.Mandelbrot,An Outline of a Purely a Phenomenological Theory of
Statistical Thermodynamics: I.Canonical Ensembles, IRE Trans.
Inform. Theory IT-2 (1956) 190; L.Rosenfeld In: Ergodic theories
Ed. by P.Caldrirola Academic Press, N.Y. (1961);
F.Schlogl,Thermodynamic Uncertainty Relation, J. Phys. Chem.
Solids 49 (1988) 679; J.Uffink and J. van Lith-van
Dis,Thermodynamic Uncertainty Relation, Found. of Phys. 29 (1999)
655
%
%
\bibitem{r13}
F.Pennini,A.Plastino, and A.R.Plastino, Power-law distributions,
Fisher information, and thermal uncertainty  [cond-mat/0110135]
%
%
\bibitem{r14}
A.E.Shalyt-Margolin and A.Ya.Tregubovich, Generalized Uncertainty
Relations in a Quantum Theory and Thermodynamics From the Uniform
Point of View [gr-qc/0204078]
%
%
\bibitem{r15}
Carlos Castro,Noncommutative  Quantum Mechanics and Geometry From
the Quantization in C-spaces  [hep-th/0206181]
%
%
\bibitem{r16}
R.P.Feynman,Statistical Mechanics,A Set of Lectures,California,
Institute of Technology.W.A.Benjamin,Inc.Advanced Book Program
Reading,Massachusets 1972
%
%
\bibitem{r17}
J.A.Wheeler,in Battele Rencontres,ed. by C.DeWitt and J.A. Wheeler
(Benjamen,NY,1968)123; B.DeWitt,Quantum Thery Gravity I.The
Canonical Theory, Phys.Rev.160(1967)1113.
%
%
\bibitem{r18}
A.H.Guth,Inflation and EternaL Inflation,[astro-ph/0002156]
%
%
\bibitem{r19}
A.E.Shalyt-Margolin and A.Ya.Tregubovich, Generalized Uncertainty
Relations in  Thermodynamics [gr-qc/0307018]
%
%
\bibitem{r20}
A.E.Shalyt-Margolin,Density Matrix in Quantum and Statistical
Mechanics at Planck-Scale [gr-qc/0307056]
%
%
\bibitem{r21}
S.F.Hassan and M.S.Martin, Trans-Plancian Effects in Inflationary
Cosmology and Modified Uncertainty Principle, [hep-th/0204110]
%
%
\end{thebibliography}
\end{document}